\documentclass[prb,preprintnumbers,amsmath,amssymb,floatfix]{revtex4}

\usepackage{graphicx}
\usepackage{subfigure}
\usepackage{dcolumn}

\begin{document}

\title{ The design of optical triode}
\author{Xiang-Yao Wu$^{a}$ \footnote{E-mail: wuxy2066@163.com},
 Ji Ma$^{a}$, Xiao-Jing Liu$^{a}$, Yu Liang$^{a}$ and Zhi-Guo Wang$^{b}$}
 \affiliation{a. Institute of Physics, Jilin Normal
University, Siping 136000 China\\
b. Department of Physics, Tongji University, Shanghai 200092,
China}

\begin{abstract}

Under the action of pump light, the conventional photonic crystal
can be turned into function photonic crystal. In the paper, we
have designed optical triode with one-dimensional function
photonic crystal, and analyzed the effect of period number, medium
thickness and refractive index, incident angle, the irradiation
way and intensity of pump light on the optical triode
magnification. We obtain some valuable results, which shall
help to optimal design optical triode. \\

\vskip 10pt


\end{abstract}

\vskip 10pt \maketitle {\bf 1. Introduction} \vskip 10pt

In 1987, E. Yablonovitch and S. John had pointed out that the
behavior of photons can be changed when propagating in the
material with periodical dielectric constant, and termed such
material Photonic Crystal [1, 2], which are designed to affect the
propagation of light [3, 4]. An important feature of the photonic
crystal is that there are allowed and forbidden ranges of
frequencies at which light propagates in the direction of index
periodicity. Due to the forbidden frequency range, known as
photonic band gap (PBG) [5, 6]. The existence of PBGs will lead to
many interesting phenomena, e.g., modification of spontaneous
emission [7-10] and photon localization [11-14]. Thus numerous
applications of photonic crystal have been proposed in improving
the performance of optoelectronic and microwave devices such as
high-efficiency semiconductor lasers, right emitting diodes, wave
guides, optical filters, high-Q resonators, antennas,
frequency-selective surface, optical limiters and amplifiers
[15-18]. These applications would be significantly enhanced if the
band structure of the photonic crystal could be tuned.

In Refs. [19-25], we have proposed function photonic crystal,
which is constituted by two media $A$ and $B$, their refractive
indexes are the functions of space position. Unlike conventional
photonic crystal (PCs), which is constituted by the constant
refractive index media $A$ and $B$. We have studied the
transmissivity and the electric field distribution with and
without defect layer.

In the paper, we have designed optical triode with one-dimensional
function photonic crystal, and analyzed the effect of period
number, medium thickness and refractive index, incident angle, the
irradiation way and intensity of pump light on the optical triode
magnification. We obtain some results: (1) When the period number,
medium thickness, incident angle and intensity of pump light
increase, the optical triode magnification increases. (2) When the
medium refractive index decrease the optical triode magnification
increases. (3) The magnification of pump light irradiation way in
Fig. 2 is larger than Fig. 3. The above results help to optimal
design optical triode.

\vskip 10pt {\bf 2. The transmissivity of one-dimensional function
photonic crystal} \vskip 10pt
\begin{figure}[tbp]
\includegraphics[width=8.5cm, height=6.5cm]{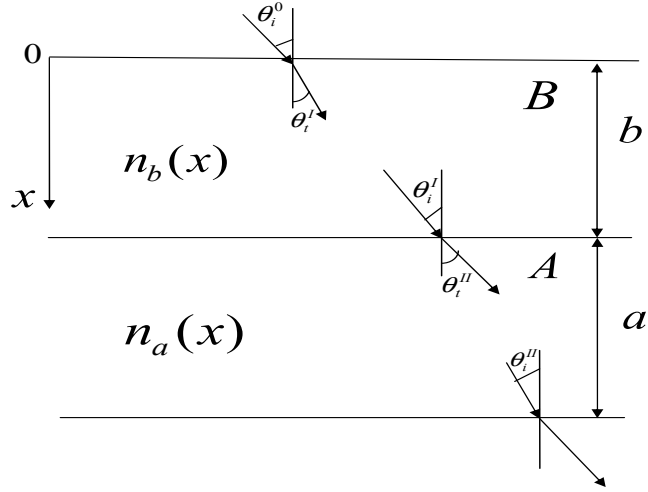}
\caption{The light transmission figure in media $B$ and $A$ of
function photonic crystal.}
\end{figure}
In Refs. [19-25], we have given the function photonic crystal
transfer matrices $M_{B}$ and $M_{A}$ of the media $B$ and $A$ for
the $TE$ wave, they are
\begin{eqnarray}
M_{B}=\left(%
\begin{array}{cc}
  \cos\delta_{b} & \frac{-i\sin\delta_{b}}{\sqrt{\frac{\varepsilon_{0}}{\mu_{0}}}n_{b}(b)\cos\theta_{i}^{I}} \\
 -in_{b}(0)\sqrt{\frac{\varepsilon_{0}}{\mu_{o}}}\cos\theta_{t}^{I}\sin\delta_{b}
 & \frac{n_{b}(0)\cos\theta_{t}^{I}\cos\delta_{b}}{n_{b}(b)\cos\theta_{i}^{I}}\\
\end{array}%
\right),
\end{eqnarray}
\begin{eqnarray}
M_{A}=\left(%
\begin{array}{cc}
 \cos\delta_{a} & -\frac{i\sin\delta_{a}}{\sqrt{\frac{\varepsilon_{0}}{\mu_{0}}}n_{a}(a)\cos\theta_{i}^{II}} \\
 -in_{a}(0)\sqrt{\frac{\varepsilon_{0}}{\mu_{o}}}\cos\theta_{t}^{II}\sin\delta_{a}
 & \frac{n_{a}(0)\cos\theta_{t}^{II}\cos\delta_{a}}{n_{a}(a)\cos\theta_{i}^{II}}\\
\end{array}%
\right),
\end{eqnarray}
where
\begin{eqnarray}
\delta_{b}=\frac{\omega}{c}n_{b}(b)\cos\theta^{I}_{i}\cdot b,
\hspace{0.2in}
\delta_{a}=\frac{\omega}{c}n_{a}(a)\cos\theta^{II}_{i}\cdot a,
\end{eqnarray}
\begin{eqnarray}
\sin\theta^{I}_{i}=\frac{n_{0}}{n_{b}(b)}\sin\theta_{i}^{0},
\hspace{0.2in} \cos\theta^{I}_{i}
=\sqrt{1-\frac{n_{0}^{2}}{n_{b}^{2}(b)}\sin^{2}\theta_{i}^{0}},
\hspace{0.2in}
\cos\theta^{I}_{t}
=\sqrt{1-\frac{n_{0}^{2}}{n_{b}^{2}(0)}\sin^{2}\theta_{i}^{0}},
\end{eqnarray}
and
\begin{eqnarray}
\sin\theta^{II}_{i}=\frac{n_{0}}{n_{a}(a)}\sin\theta_{i}^{0},
\hspace{0.2in} \cos\theta^{II}_{i}
=\sqrt{1-\frac{n_{0}^{2}}{n_{a}^{2}(a)}\sin^{2}\theta_{i}^{0}},
\hspace{0.2in}
\cos\theta^{II}_{t}
=\sqrt{1-\frac{n_{0}^{2}}{n_{a}^{2}(0)}\sin^{2}\theta_{i}^{0}}.
\end{eqnarray}
In one period, the transfer matrix $M$ is
\begin{eqnarray}
&&M=M_{B}\cdot M_{A}\nonumber\\
&&=\left(%
\begin{array}{cc}
  \cos\delta_{b} & \frac{-i\sin\delta_{b}}{\sqrt{\frac{\varepsilon_{0}}{\mu_{0}}}n_{b}(b)\cos\theta_{i}^{I}} \\
 -in_{b}(0)\sqrt{\frac{\varepsilon_{0}}{\mu_{o}}}\cos\theta_{t}^{I}\sin\delta_{b}
 & \frac{n_{b}(0)\cos\theta_{t}^{I}\cos\delta_{b}}{n_{b}(b)\cos\theta_{i}^{I}}\\
\end{array}%
\right) \nonumber\\&&
\left(%
\begin{array}{cc}
   \cos\delta_{a} & \frac{-i\sin\delta_{a}}{\sqrt{\frac{\varepsilon_{0}}{\mu_{0}}}n_{a}(a)\cos\theta_{i}^{II}} \\
 -in_{a}(0)\sqrt{\frac{\varepsilon_{0}}{\mu_{o}}}\cos\theta_{t}^{II}\sin\delta_{a}
 &\frac{n_{a}(0)\cos\theta_{t}^{II}\cos\delta_{a}}{n_{a}(a)\cos\theta_{i}^{II}}\\
\end{array}%
\right).
\end{eqnarray}
Where $n_{b}(0)$, $n_{b}(b)$, $n_{a}(0)$ and $n_{a}(a)$ are the
starting point and endpoint values of refractive indices for media
$B$ and $A$, $b$ and $a$ are the thickness of media $B$ and $A$,
$\theta_{i}^{0}$ is incident angle, $n_{0}$ is air refractive
index, and the angles $\theta_{t}^{I}$, $\theta_{i}^{I}$,
$\theta_{t}^{II}$ and $\theta_{i}^{II}$ are shown in FIG. 1.

We can find the transfer matrix $M$ of the function photonic
crystal is more complex than the conventional PCs. For the
structure $(BA)^N$ function photonic crystal, its characteristic
equation is
\begin{eqnarray}
\left(%
\begin{array}{c}
  E_{1} \\
  H_{1} \\
\end{array}%
\right)&=&M_{B}M_{A}M_{B}M_{A}\cdot\cdot\cdot M_{B}M_{A}\left(%
\begin{array}{c}
  E_{N+1} \\
  H_{N+1} \\
\end{array}%
\right)
\nonumber\\&=&M\left(%
\begin{array}{c}
  E_{N+1} \\
  H_{N+1} \\
\end{array}%
\right)=\left(%
\begin{array}{c c}
  A &  B \\
 C &  D \\
\end{array}%
\right)
 \left(%
\begin{array}{c}
  E_{N+1} \\
  H_{N+1} \\
\end{array}%
\right).
\end{eqnarray}
With the total transfer matrix $M$, we can obtain the transmission
coefficient $t$, it is
\begin{eqnarray}
t=\frac{E_{N+1}}{E_{in}}=\frac{E_{out}}{E_{in}}=\frac{2\eta_{0}}{A\eta_{0}+B\eta_{0}\eta_{N+1}+C+D\eta_{N+1}},
\end{eqnarray}
where $E_1=E_{in}+E_{r}$, $E_{in}$ is the incident electric field,
$E_{r}$ is the reflected electric field, $E_{N+1}=E_{out}$ is the
output electric field and
$\eta_{0}=\eta_{N+1}=\sqrt{\frac{\varepsilon_0}{\mu_0}}\cos\theta
_{i}^0 $.

 \vskip 10pt {\bf 2. The design principle of optical triode} \vskip
10pt

In Refs. [20-25], we can find the transmissivity can be larger
than $1$ and the output electric field intensity has been
magnified for the function photonic crystal that the refractive
indexes of media $B$ and $A$ are the increasing functions with the
space position. In Refs. 23 and 25, we have designed and optimally
designed optical device, such as optical amplifier, attenuator,
and optical diode by the function photonic crystal.

In the following, we shall explain how to turn the conventional
photonic crystal into the function photonic crystal, and give the
design principle of optical triode.
\begin{figure}[tbp]
\includegraphics[width=10cm, height=5cm]{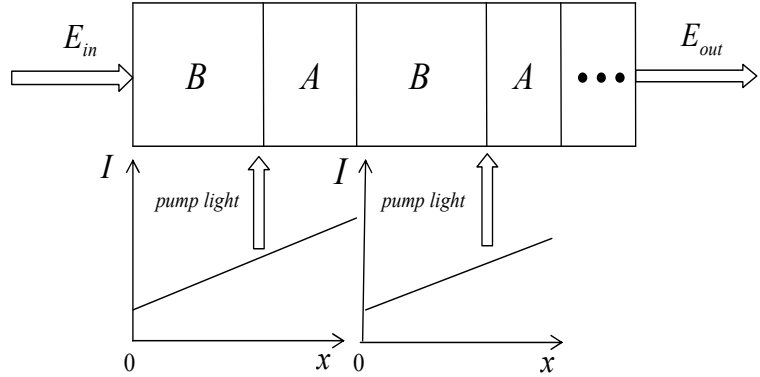}
\caption{The pump light irradiate vertically photonic crystal, its
one-period irradiate medium $(BA)$.}
\end{figure}
In nonlinear optics, the medium refractive index is the linear
function of light intensity $I$, which is called the optical kerr
effect, it is
\begin{eqnarray}
n(I)=n_0+n_{2}I,
\end{eqnarray}
where $n_0$ represents the usual, weak-field refractive index, the
optical Kerr coefficient $n_{2}=\frac{3}{4n_0^2\epsilon_0
c}\chi^{(3)}$, and $\chi^{(3)}$ is the third-order nonlinear
optical susceptibility.

In Fig. 2, in the perpendicular to the direction of incident
light, we join the strong laser field (pump light) to every
one-period $BA$ of conventional photonic crystal respectively, the
intensity $I$ distribution is the function of space position.
Here, the pump light intensity $I$ is the linear function of every
one-period $BA$ thickness $x$ and width $y$, it is
\begin{eqnarray}
I=I_0(x\times y),
\end{eqnarray}
the intensity distribution $I$ is evenly in the $y$ direction, we
can let $y=1$.

Substituting Eq. (10) into (9), there is
\begin{eqnarray}
n(x)=n_0+n_{2}I_0x,
\end{eqnarray}
where $I_0$ is the intensity coefficient of pump light, the
optical kerr effect of incident light is neglected. The
conventional medium refractive index $n_0$ become the linear
function $n(x)$ of space position $x$, i.e., the refractive
indices $n_b$ and $n_a$ of conventional media $B$ and $A$ become
the linear functions $n_b(x)$ and $n_a(x)$. The Fig. 2 has turned
the conventional photonic crystal into the function photonic
crystal with the pump light. At every one-period $BA$, the
refractive indexes starting point and endpoint value of media $B$
and $A$ are same, they are
\begin{eqnarray}
n_{b}(0)=n_{b}, \hspace{0.2in} n_{b}(b)=n_{b}+n_2 I_{0}b,
\end{eqnarray}
\begin{eqnarray}
n_{a}(0)=n_{a}+n_2 I_{0}b, \hspace{0.2in} n_{a}(a)=n_{a}+n_2
I_{0}(a+b),
\end{eqnarray}
where $n_{b} (n_{a})$ is the refractive index of conventional
medium $B (A)$ (without joining pump light), $n_{b}(0)
(n_{a}(0))$, $n_{b}(b)(n_{a}(a))$ are the refractive indices
starting point and endpoint value of function medium $B(A)$ (with
joining pump light), and $b (a)$ is the thickness of media $B
(A)$. In every one-period, the refractive indices of media $B$ and
$A$ are the linear functions of space position $x$.
\begin{figure}[tbp]
\includegraphics[width=10cm, height=5cm]{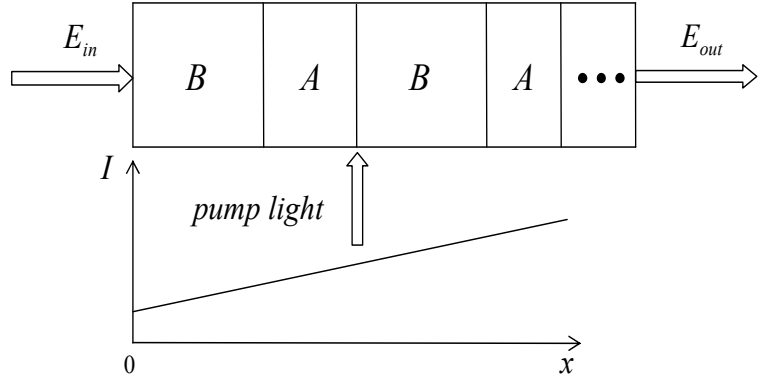}
\caption{The pump light irradiate vertically photonic crystal, its
one-period irradiate medium $(BA)^2$.}
\end{figure}

In Fig. 3, we join the pump light to photonic crystal every
two-period $(BA)^2$, respectively. The pump light intensity $I$ is
the linear function of every two-period $(BA)^2$ thickness $x$. At
every two-period, the refractive indexes of media $B$ and $A$ are
same, they are
\begin{eqnarray}
n_{1b}(0)=n_{b}, \hspace{0.2in} n_{1b}(b)=n_{b}+n_2 I_{0}b,
\end{eqnarray}
\begin{eqnarray}
n_{1a}(0)=n_{a}+n_2 I_{0}b, \hspace{0.2in} n_{1a}(a)=n_{a}+n_2
I_{0}(a+b),
\end{eqnarray}
\begin{eqnarray}
n_{2b}(0)=n_{b}+n_2 I_{0}(b+a), \hspace{0.2in} n_{2b}(b)=n_{b}+n_2
I_{0}(2b+a),
\end{eqnarray}
\begin{eqnarray}
n_{2a}(0)=n_{a}+n_2 I_{0}(2b+a), \hspace{0.2in}
n_{2a}(a)=n_{a}+n_2 I_{0}(2b+2a),
\end{eqnarray}
where $n_{1b}(0) (n_{1a}(0))$, $n_{1b}(b)(n_{1a}(a))$ are the
refractive index starting point and endpoint value of the first
period medium $B(A)$, and $n_{2b}(0) (n_{2a}(0))$,
$n_{2b}(b)(n_{2a}(a))$ are the refractive index starting point and
endpoint value of the second period medium $B(A)$. In every
two-period, the refractive indices of media $B$ and $A$ are linear
functions of space position $x$.

The function photonic crystal Figs. 2 and 3 should be designed
optical triode, the optical triode magnification $\beta$ is
defined as
\begin{eqnarray}
\beta=|t|=|\frac{E_{out}}{E_{in}}|=|\frac{2\eta_{0}}{A\eta_{0}+B\eta_{0}\eta_{N+1}+C+D\eta_{N+1}}|.
\end{eqnarray}

\begin{figure}[tbp]
\includegraphics[width=8.5 cm, height=8cm]{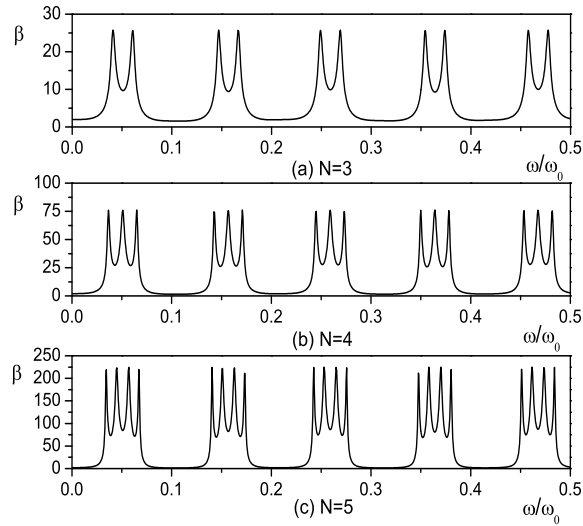}
\caption{The effect of period numbers $N$ on magnification $\beta$
under the pump light action of Fig. 2. (a) $N=3$, (b) $N=4$ and
(c) $N=5$.}
\end{figure}

\vskip 8pt {\bf 3. Numerical result} \vskip 8pt

In this section, we report our numerical results of the optical
triode magnification $\beta$ for the one-dimensional function
photonic crystal. The main parameters are: The media $B$ and $A$
thickness $b=398nm$, $a=208nm$, and the conventional media $B$ and
$A$ (without joining pump light) refractive indices $n_{b}=1.68$,
$n_{a}=2.56$, the incident angle $\theta_{i}^{0}=0$, the central
frequency $\omega_0=1.216\times10^{15}Hz$, the optical Kerr
coefficient $n_{2}$ in the range of
$9.0\times10^{-17}\sim2.3\times10^{-6} (cm^2/W)$, we take
$n_{2}=10^{-6}(cm^2/W)$ and the light intensity coefficient
$I_0=2.2\times10^{9}(cm^2/W)$. Firstly, we calculate the optical
triode magnification $\beta$ under the pump light action of Fig.
2. Substituting Eqs. (12) and (13) into (6), we can obtain the
transfer matrix $M$ of one period, with Eqs. (7), (8) and (18), we
can calculate the magnification $\beta$, they are shown in Figs.
4-9, which give the relation between magnification $\beta$ and
incident light frequency $\omega$. In Fig. 4 (a), (b) and (c), we
give the magnification $\beta$ corresponding to period numbers
$N=3$, $N=4$ and $N=5$, respectively. From Fig. 4, we can obtain
some results: (1) The magnification $\beta$ is related to the
incident light frequency, the incident electric field of certain
frequencies get through the function photonic crystal, the output
electric field have been magnified obviously. (2) When period
numbers $N=3$, $N=4$ and $N=5$, the magnifications amplitude
$\beta_{max}$ are $25$, $75$ and $225$, i.e., the output electric
field intensity is $\beta_{max}$ multiple of input electric field
intensity, which is shown in Fig. 5. The Fig. 5 (a) and (b) are
input electric field and output electric field, respectively. The
output electric field has been magnified. In Fig. 6 (a) and (b),
the media $B$ thicknesses are $b=398nm$ and $498 nm$, we can find
the magnification $\beta$ increases with media $B$ thickness
increasing. In Fig. 7 (a) and (b), the media $B$ refractive
indices are $n_b=1.68$ and $n_b=1.98$, we can find the
magnification $\beta$ decreases with media $B$ refractive index
increasing. In Fig. 8 (a) and (b),  the intensity coefficients
$I_0$ of pump light are $I_0=2.2\times10^{9}(cm^2/W)$ and
$I_0=3.2\times10^{9}(cm^2/W)$, we can find the magnifications
amplitude $\beta_{max}$ increases with the intensity coefficients
increasing. In Fig. 9 (a), (b) and (c), we give the magnification
$\beta$ corresponding to incident angles
$\theta_i^0=\frac{\pi}{12}$, $\theta_i^0=\frac{\pi}{6}$ and
$\theta_i^0=\frac{\pi}{3}$, respectively. From Fig. 9, we can
obtain some results: (1) The magnifications amplitude
$\beta_{max}$ increases with incident angle increasing. (2) When
incident angle $\theta_i^0=\frac{\pi}{12}$,
$\theta_i^0=\frac{\pi}{6}$ and $\theta_i^0=\frac{\pi}{3}$, the
$\beta_{max}$ are $80$, $90$ and $120$. Nextly, we calculate the
optical triode magnification $\beta$ under the pump light action
of Fig. 3. Substituting Eqs. (14) to (17) into (1) and (2), we can
obtain the transfer matrices $M_{B1}$, $M_{A1}$, $M_{B2}$ and
$M_{A2}$, and the transfer matrix of one period is
$M=M_{B1}M_{A1}M_{B2}M_{A2}$. With Eqs. (7), (8) and (18), we can
calculate the magnification $\beta$, which is shown in Fig. 10. In
Fig. 10 (a), (b) and (c), we give the magnification $\beta$
corresponding to period numbers $N=4$, $N=6$ and $N=8$,
respectively. From Fig. 10, we can obtain some results: (1) The
magnifications amplitude $\beta_{max}$ increases with period
numbers increasing, when period numbers $N=4$, $N=6$ and $N=8$,
the $\beta_{max}$ are $15$, $70$ and $250$. (2) For the different
irradiation way of pump light (Figs. 2 and 3), the $\beta-\omega$
distribution and the magnifications amplitude $\beta_{max}$ are
different. Obviously, the irradiation way of Fig. 2 can obtain the
more magnification.

The reason of optical triode realizing the light amplification is
the incident light absorb the pump light energy. We can compare
the triode triode with electronic triode. The incident and output
electronic fields are the equal of the input and output
alternating electrical signals, and the pump light amount to
direct-current power.

\begin{figure}[tbp]
\includegraphics[width=8.5 cm, height=8cm]{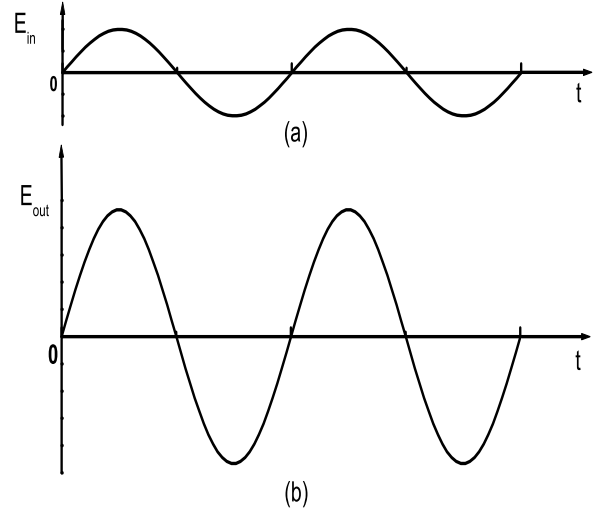}
\caption{The input and output electric field of photonic crystal.
(a) The input electric field $E_{in}$. (b) The output electric
field $E_{out}$ has been amplified, and there are $E_{out}=\beta
E_{in}$.}
\end{figure}

\begin{figure}[tbp]
\includegraphics[width=8.5 cm, height=8cm]{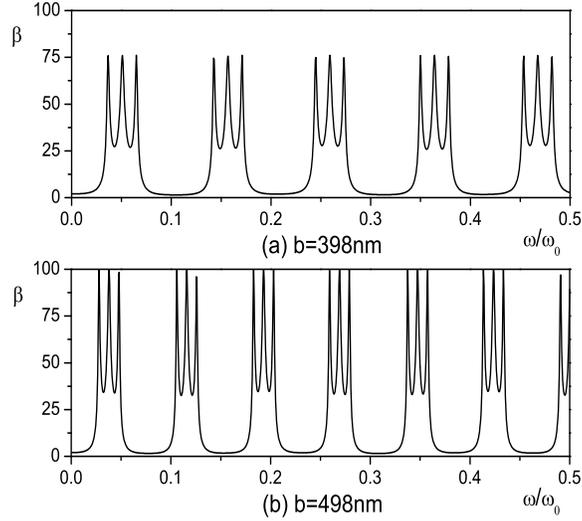}
\caption{The effect of medium $B$ thickness on magnification
$\beta$ under the pump light action of Fig. 2. (a) $b=398nm$ and
(b) $b=498nm$.}
\end{figure}

\newpage
\vskip 10pt {\bf 4. Conclusion} \vskip 10pt

In summary, we have designed optical triode with one-dimensional
function photonic crystal. We analyzed the effect of period
number, medium thickness and refractive index, incident angle, the
irradiation way and intensity of pump light on the optical triode
magnification $\beta$, and obtained some results: (1) When the
period number, medium thickness, incident angle and intensity of
pump light increase, the optical triode magnification increases.
(2) When the medium refractive index decrease the optical triode
magnification increases. (3) The magnification of pump light
irradiation way in Fig. 2 is larger than Fig. 3. The above results
help to optimal design optical triode.

\vskip 12pt {\bf 5.  Acknowledgment} \vskip 12pt

This work was supported by the National Natural Science Foundation
of China (no.61275047), the Research Project of Chinese Ministry
of Education (no.213009A) and Scientific and Technological
Development Foundation of Jilin Province (no.20130101031JC).

\newpage
\begin{figure}[tbp]
\includegraphics[width=8.5cm, height=8cm]{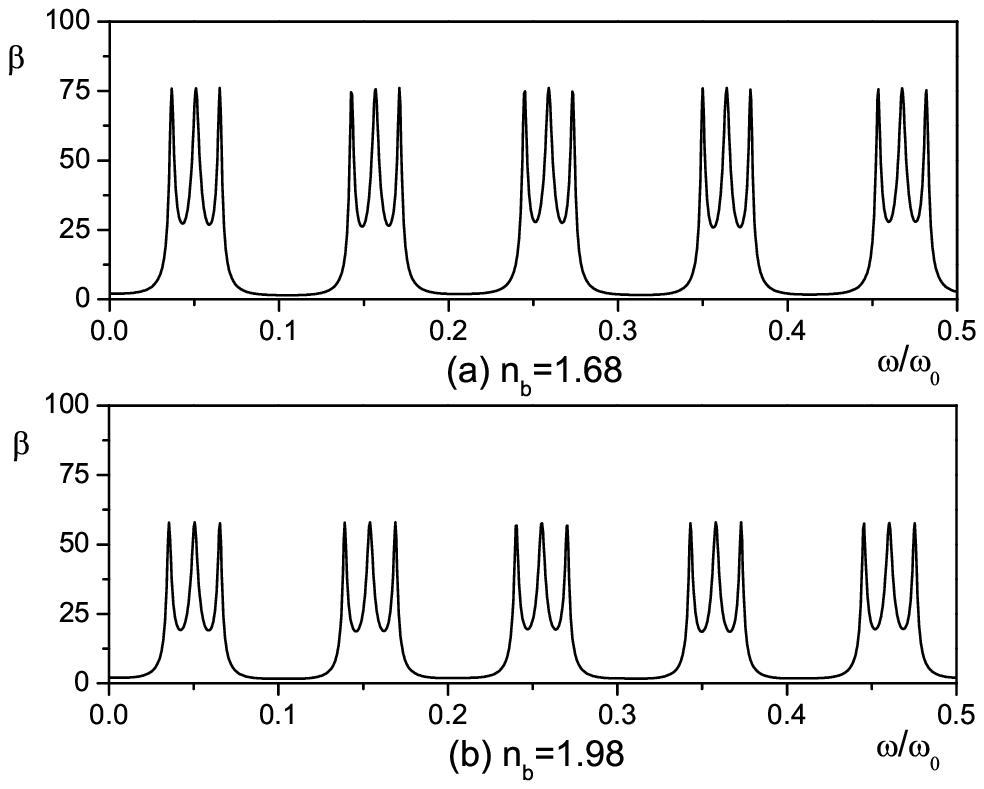}
\caption{The effect of medium $B$ refractive index on
magnification $\beta$ under the pump light action of Fig. 2. (a)
$n_b=1.68$ and (b) $n_b=1.98$.}
\end{figure}
\begin{figure}[tbp]
\includegraphics[width=8.5 cm, height=8cm]{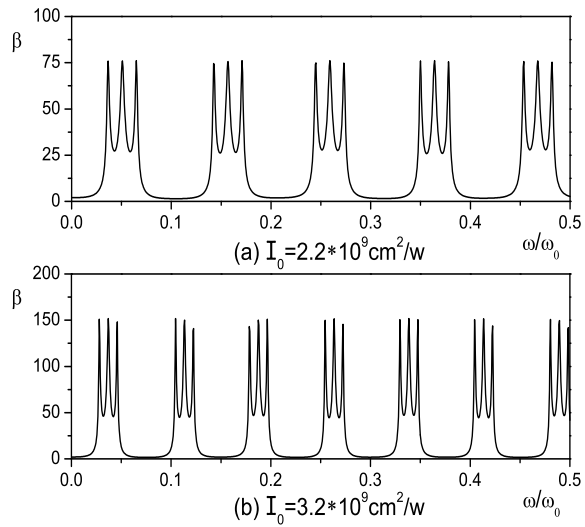}
\caption{The effect of the intensity coefficient $I_0$ on
magnification $\beta$ under the pump light action of Fig. 2. (a)
$I_0=2.2\times10^{9}(cm^2/W)$ and (b)
$I_0=3.2\times10^{9}(cm^2/W)$.}
\end{figure}

\begin{figure}[tbp]
\includegraphics[width=8.5 cm, height=8cm]{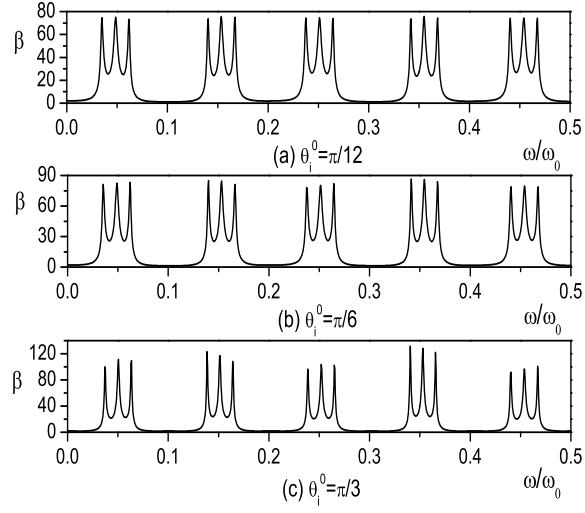}
\caption{The effect of the incident angle $\theta_i^0$ on
magnification $\beta$ under the pump light action of Fig. 2. (a)
$\theta_i^0=\frac{\pi}{12}$, (b) $\theta_i^0=\frac{\pi}{6}$ and
(c) $\theta_i^0=\frac{\pi}{3}$.}
\end{figure}

\begin{figure}[tbp]
\includegraphics[width=8.5 cm, height=8cm]{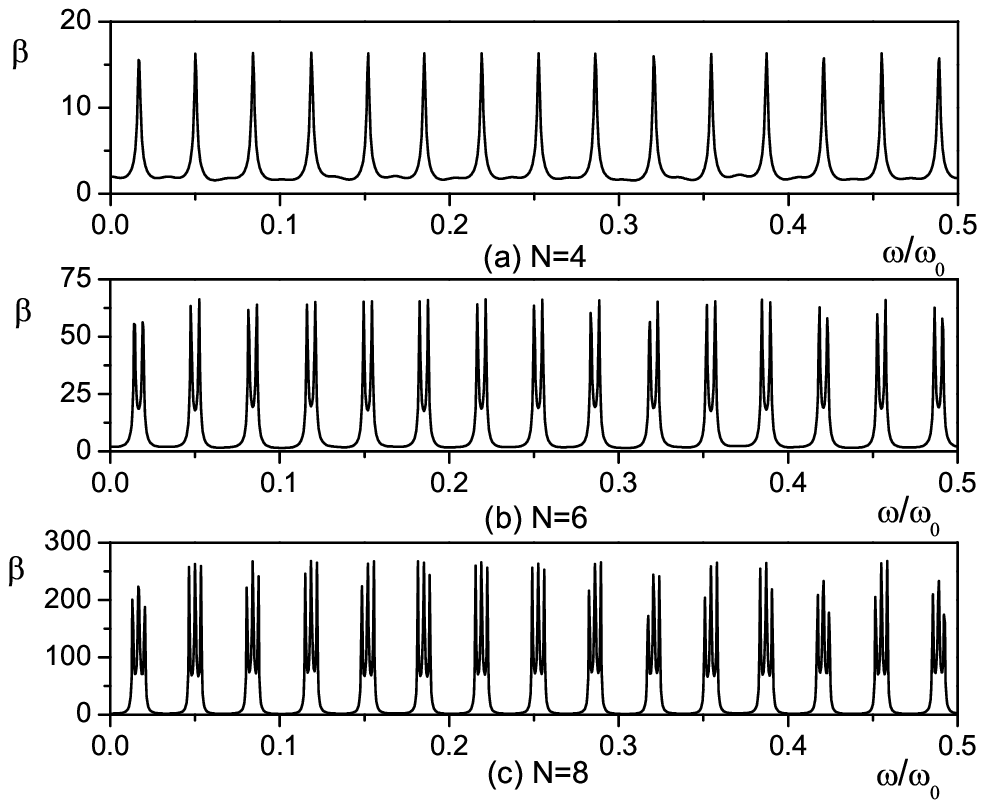}
\caption{The effect of period numbers $N$ on magnification $\beta$
under the pump light action of Fig. 3. (a) $N=4$, (b) $N=6$ and
(c) $N=8$.}
\end{figure}
\end{document}